\documentclass[aps,prd,preprintnumbers,nofootinbib,preprint]{revtex4}
\usepackage[utf8]{inputenc}
\usepackage{bm}
\usepackage{latexsym}
\usepackage{dcolumn}
\usepackage{amsmath,amsfonts,amssymb}
\usepackage{graphicx,epsfig}
\usepackage{soul}
\usepackage{ulem}
\usepackage{amsthm}
\usepackage{color}

\begin{document}
\title{Is Gravity Actually the Curvature of Spacetime?}
\author{{Sebastian Bahamonde}$^{1,2}$}\email[email:~]{sbahamonde@ut.ee, sebastian.beltran.14@ucl.ac.uk}
\author{Mir~Faizal$^{^{3,4}}$} \email[email:~]{mirfaizalmir@googlemail.com}
%
%
\affiliation{$^1$Laboratory of Theoretical Physics, Institute of Physics, University of Tartu, W. Ostwaldi 1, 50411 Tartu, Estonia.}
\affiliation{$^2$Department of Mathematics, University College London, Gower Street, London, WC1E 6BT, United Kingdom.}
\affiliation{$^3$Department of Physics and Astronomy, University of Lethbridge, 4401 University Drive, Lethbridge, Alberta T1K 3M4, Canada}
\affiliation{$^4$Irving K. Barber School of Arts and Sciences, University of British Columbia - Okanagan, 3333 University Way, Kelowna, British Columbia V1V 1V7, Canada}

%



 \begin{abstract}
The  Einstein   equations, apart from being the classical field equations of General Relativity, are also 
the classical field equations of two other theories of gravity. 
As  the experimental tests of General Relativity are done using  the Einstein  equations,   we do not really know, if gravity is the curvature of a torsionless spacetime, 
or torsion of a curvatureless spacetime, or if it  occurs due to the non-metricity of a curvatureless and torsionless spacetime.
However, as the classical actions 
of all these theories differ from each other by  boundary terms, and the Casimir effect is a boundary effect, 
we  propose that a novel gravitational 
Casimir effect between superconductors can be used to test  which of these theories actually  describe  gravity. 
\\
{}
\\
{\underline{Corresponding author:}~~Sebastian Bahamonde}~~~~~~~
{\underline{Submission date:}~~31 March 2019}\\
{}\\
{\bf Essay written for the Gravity Research Foundation 2019 Awards for Essays on Gravitation}
\end{abstract}

\maketitle
\textit{General Relativity} (GR)  is one of the most well tested theories in Nature, but in all those tests, what is actually tested 
are  the predictions made by the  Einstein   equations   \cite{1}.  It is possible to construct  two other  geometrical theories describing  gravity, 
which are fundamentally different from
GR, but  whose classical field equations are the  Einstein   equations. To understand these theories, we first note that the spacetime has to  be  described by  a differential manifold   in any geometrical theory of gravity. 
Now a general affine connection  $\Gamma^{\alpha}{}_{\mu\nu}$, on such a manifold,  can be decomposed into
three pieces~\cite{BeltranJimenez:2017tkd, BeltranJimenez:2018vdo}
	\begin{equation}
	\Gamma^{\alpha}{}_{\mu\nu}=\left\{^{\phantom{i} \alpha}_{\mu\nu}\right\} +K^{\alpha}{}_{\mu\nu}+L^{\alpha}{}_{\mu\nu}\,.\label{Eq1}
	\end{equation}
 The first term $\left\{^{\phantom{i} \alpha}_{\mu\nu}\right\}$ is the standard \textit{Levi-Civita connection}, which
is obtained from the metric. The second term $K^{\alpha}{}_{\mu\nu}$ is the  \textit{contortion tensor}, 
 which is obtained from the torsion tensor $T^{\alpha}{}_{\mu\nu}$ as  $K^{\alpha}{}_{\mu\nu}\equiv (1/2) T^{\alpha}{}_{\mu\nu}+T_{(\mu}{}^{\alpha}{}_{\nu)}$.
 The last term  $L^{\alpha}{}_{\mu\nu}$ is the
 \textit{disformation tensor}, which is constructed from the non-metricity tensor $Q_{\alpha\mu\nu}\equiv \nabla_{\alpha}g_{\mu\nu}$
 as $L^{\alpha}{}_{\mu\nu}\equiv (1/2) Q^{\alpha}{}_{\mu\nu}-Q_{(\mu}{}^{\alpha}{}_{\nu)}$.
	
{   GR is described by a torsionless spacetime   ($T^{\alpha}{}_{\mu\nu}=0$), which   satisfies the 
	 metric compatibility condition $\nabla_{\alpha}g_{\mu\nu}=0$ ($Q_{\alpha\mu\nu}=0$). So, as 
	 $K^{\alpha}{}_{\mu\nu}=L^{\alpha}{}_{\mu\nu}=0$ in GR, the affine connection of Eq.~\eqref{Eq1} can be written in terms of   the Levi-Civita connection as
	 $\Gamma^{\alpha}{}_{\mu\nu}=\left\{^{\phantom{i} \alpha}_{\mu\nu}\right\}$. The   curvature tensor  constructed from this Levi-Civita connection $\bar{{R}}^\alpha_{\phantom{\alpha}\beta\mu\nu} $ is used to obtain the Einstein-Hilbert action  
 \begin{equation}
 \mathcal{S}_{\rm G}=\frac{1}{16\pi G}\int \sqrt{-g} \bar{R}\,,\label{1}
 \end{equation}
where $g$ is the determinant of the metric $g_{\mu\nu}$ and $\bar{R}\equiv g^{\beta\nu}\bar{R}{}^\alpha{}_{\beta\alpha\nu}$ is the scalar curvature obtained  from $\bar{{R}}^\alpha_{\phantom{\alpha}\beta\mu\nu} $. Einstein   equations are the classical field equations obtained  from this action. 
  }

{\textit{Teleparallel Gravity} (TG) is another geometrical theory of gravity, whose classical field equations are 
the Einstein equations. In this theory,  the general connection of   Eq.~\eqref{Eq1}  is equated to the  \textit{Weitzenb\"{o}ck connection}, and  so the curvature of spacetime vanishes. Thus, TG is constructed using such a  curvatureless spacetime, which
satisfies the metric  compatibility condition $Q_{\alpha\mu\nu}=0$ ($L^{\alpha}{}_{\mu\nu}=0$) ~\cite{tele, ft14, tele12, tele14, tele15, tele16}.
  This theory is constructed from  the torsion tensor in the tetrad formalism, and its action is given by 
 \begin{equation}
\mathcal{S}_{\rm T}=-\frac{1}{16\pi G}\int e T\,,\label{2}
\end{equation}
where $e=\sqrt{-g}$ is the determinant of the tetrad and $T$ is the scalar torsion (which is constructed from  contractions of the torsion tensor).

It may be noted that in curvatureless spacetime of TG,  the curvature tensor $({R}^{\alpha}{}_{\beta\mu\nu})$  obtained from the general affine connection of  Eq.~\eqref{Eq1} vanishes, but the curvature tensor constructed using the Levi-Civita connection $(\bar{R}^{\alpha}{}_{\beta\mu\nu})$ does not vanish. In TG, the scalar curvature $R$ obtained from ${R}^{\alpha}{}_{\beta\mu\nu}$  is related to the scalar curvature $\bar{R}$ obtained from  $\bar{R}^{\alpha}{}_{\beta\mu\nu}$, as $R = \bar{R} + T -(2/e)\partial_\mu (e T^{\lambda}{}_{\lambda}{}^\mu)=0$, so we can write
\begin{align}
\bar{R}=-T+  B_{\rm T}\,, 
\end{align}
where $B_{\rm T} = (2/e)\partial_\mu (e T^{\lambda}{}_{\lambda}{}^\mu)$  is a boundary term. Thus,  the
 action for GR given by Eq. \eqref{1} and the action  for TG given by Eq. \eqref{2}, differ from each other by the boundary term $B_{\rm T}$. }

 { It is also possible to   formulate a 
  \textit{Theory of  Non-Metricity}   (TNM) to describe gravity \cite{sy, sy12, sy14, sy15, sy16}.  This theory is also called Coincident General Relativity or Symmetric Teleparallel Gravity, as it has certain features which resemble both GR and TG,  but we shall call it as TNM, as the theory is based on the  concept of non-metricity. 
  In this  theory, both  the torsion tensor  
  and 
  $R^{\lambda}{}_{\mu\nu\beta}$  vanish, and  gravity is produced because of the non-vanishing  non-metricity tensor,  $\nabla_{\alpha}g_{\mu\nu}=Q_{\alpha\mu\nu}\neq 0 $ ($L_{\alpha}{}_{\mu\nu}\neq 0 $).
 The action for this theory is constructed  using   the non-metricity scalar $Q$ (which is obtained from the non-metricity tensor $Q_{\alpha\mu\nu}$) as 
\begin{equation}
\mathcal{S}_{\rm N}=-\frac{1}{16\pi G}\int \sqrt{-g} Q\,.\label{3}
\end{equation}
As TNM is described by a torsionless  and curvatureless  spacetime, $Q$ can be related  to $\bar{R}$ (curvature obtained from the Levi-Civita connection) as 
\begin{equation}
\bar{R}=-Q+ B_{\rm N}\,, 
\end{equation}
where  $B_{\rm N} = ({1}/{\sqrt{-g}})\partial_{\alpha}(Q^{\alpha}{}_{\lambda}{}^{\lambda}-Q^{\lambda}{}_{\lambda}{}^{\alpha})$ is again a boundary  term (different from the  boundary term  obtained in TG). So, the   action for GR given by Eq.~\eqref{1} and the action for TNM given by  Eq.~\eqref{3}  differ from each other by  the boundary term $B_{\rm N}$.     }

Even though  the actions of GR, TG and TNM differ from each other  by boundary terms, they  have the same classical field equations (Einstein equations), so they cannot  be  classically     distinguish from each other. The only reason for the  preferential attention given to GR (over the other two geometrical theories) is  historical and not scientific.
{However, they can be differentiated using quantum effects because    these theories are 
fundamentally different from each other and  will  produce different quantum corrections. 
We do not have a full 
theory of quantum gravity, but it  is possible to get an estimate of perturbative quantum gravitational effects,
using the formalism of effective field theories \cite{effective, effectivegr, effectivetp}.    Thus, the classical 
  actions for GR ($\mathcal{S}_{\rm G}$),    
   TG ($\mathcal{S}_{\rm T}$) and  TNM ($\mathcal{S}_{\rm N}$) get
  corrected by   quantum corrections $\mathcal{S}_{\rm QG}$,
$\mathcal{S}_{\rm QT}$ and  $\mathcal{S}_{\rm QN}$, such that
\begin{eqnarray}
\mathcal{S}_1 = \mathcal{S}_{\rm G} + \mathcal{S}_{\rm QG}\,,\quad
\mathcal{S}_2 = \mathcal{S}_{\rm T} + \mathcal{S}_{\rm QT}\,,\quad \mathcal{S}_3 = \mathcal{S}_{\rm N} + \mathcal{S}_{\rm QN}\,. 
\end{eqnarray}
It is not possible to use cosmological and astrophysical observations to differentiate between  $\mathcal{S}_{\rm G}$, $\mathcal{S}_{\rm T}$ and $\mathcal{S}_{\rm N}$, however,  such observations can differentiate between $\mathcal{S}_1$,  $\mathcal{S}_2$ and $\mathcal{S}_3$. It has been demonstrated that the    quantum corrected GR \cite{effectivegr}   and quantum corrected  TG 
\cite{effectivetp} are both consistent
with the    cosmological data obtained 
from SNe Ia + BAO + CC + $H_0$  \cite{nonl1, nonl2, nontp1, nontp2}, and so at present, quantum corrections cannot rule out either of them.     However, it  is still  possible that   future cosmological observations may  rule out one of these theories.}

Even though, at present, we are not able to use quantum corrections  to differentiate  between these theories, it is still   possible to use 
a combination of quantum effects and boundary effects to distinguish  them from each other. As the actions of GR, TG, TNM differ from each other by boundary terms, and 
the  Casimir effect is a quantum mechanical boundary effect, a gravitational Casimir effect can be used to  distinguish   them from each other. The reflection of  gravitational waves in the microwave regime by quantum properties of superconductors (Heisenberg-Coulomb  effect)~\cite{Minter2010, mini, mini2, mini4} can produce a novel measurable   gravitational Casimir effect  
\cite{ca, ca12, ca14, ca15, mini4}. In ordinary metal plates, the lattice of ions and electrons move 
along the same geodesic, in the presence of gravitational waves. However, when  Cooper pairs  form below the superconducting transition, they    move along a  non-geodesic path due to their quantum
non-localizability.   
It has been demonstrated that this  produces a large mass conductivity  due to  an enhanced mass current~\cite{ca, ca12, ca14, ca15, mini4}. As   the electromagnetic waves are  reflected due to the electrical conductivity, this mass conductivity   reflections 
gravitational waves~\cite{Minter2010, mini, mini2, mini4}.
Thus, for such systems,   a  gravitational Casimir effect can be produced   \cite{ca, ca12, ca14, ca15, mini4}, in analogy with the  conventional electromagnetic Casimir effect \cite{em12,em15,em16, em14}.

As the actions for GR, TG and TNM are related to each other by boundary terms, we can relate  the gravitational Casimir energy in  GR   $(\langle E \rangle _{G})$ \cite{ca, ca12, ca14, ca15, mini4}   to the gravitational Casimir energies in TG $(\langle E \rangle _{T})$ and TNM $(\langle E \rangle _{N})$ 
as  
\begin{equation}
    \langle E \rangle _T = \langle E \rangle _{G} +  \langle E \rangle _{B_{\rm T}}\,,\quad 
\langle E \rangle _{N} = \langle E \rangle _{G} +  \langle E \rangle _{B_{\rm N}}\,,
\end{equation}
where $ \langle E \rangle _{B_{\rm T}}$ is the contribution from the  boundary term ${B_{\rm T}}$, and $\langle E \rangle _{B_{\rm N}}$ is the contribution from   boundary term ${B_{\rm N}}$. Since the boundary action for these theories is different,  so  $\langle E \rangle _{{B_{\rm T}}}\neq \langle E \rangle _{{B_{\rm N}}}\neq 0$, thus we obtain 
 $\langle E \rangle _{G} \neq\langle E \rangle _{T}\neq \langle E \rangle _{{B_{\rm N}}}$. So,   these theories will produce different   gravitational Casimir effects, and such  effects can be used to  test which of these  theories is actually the geometrical theory of  gravity.  
It may be noted that the      Casimir force between superconductors has been recently  experimentally  measured  \cite{mini4, 
mini01, mini02, mini04, mini05, mini07}. 
Thus, it is possible to measure   the novel  gravitational Casimir effect due to the onset of superconductivity between two aluminum nanostrings. With an
optomechanical cavity readout, these experiments could detect  $6~mPa$ differences in the Casimir force between such nanostrings  \cite{mini4, 
mini01, mini02, mini04, mini05, mini07}.  The magnitude of  a gravitational  Casimir effect  depends on the difference between the   change in momentum  of the Cooper pair and change in the momentum of the ion core   \cite{ca, ca12, ca14, ca15, mini4}. Even if a more detailed analysis, 
reduced the magnitude of this novel gravitational Casimir effect by ten orders of magnitude, it would still remain a measurable  effect, using the currently available technology. It is important to achieve   sufficiently accurate  parallelism between two superconductors  at low temperatures  to produce this novel gravitational  Casimir effect. The technology needed to obtain such an accurate parallelism has  already been used in  resonator platforms for superconducting 
circuits \cite{lc, lc12}. So, such an  experiment can be performed using the currently available  technology, and  we can know which theory actually describes gravity in our Universe. 

\section*{Acknowledgments}
The authors thank Richard Norte  and Simon Groblacher for the helping us  develop the argument for measuring  gravitational  Casimir effect. The authors also thank Konstantinos F. Dialektopoulos for the feedback provided to improve the essay. S.B. is supported by Mobilitas Pluss N$^{\circ}$ MOBJD423 by the Estonian government

\end{document}